# High Speed CAN Transmission Scheme Supporting Data Rate of over 100 Mbps

Suwon Kang, Ph.D, IEEE Member, Sungmin Han, Seungik Cho, Donghyuk Jang, Hyuk Choi, and Ji-Woong Choi, DGIST, IEEE Senior Member

*As the number of electronic components in the car increases, the requirement for the higher data transmission scheme among them is on the sharp rise. Controller area network (CAN) has been widely adopted to support the in-car communications needs but the data rate is far below what other schemes such as Ethernet and optical fibers can offer. A new scheme for enhancing the speed of CAN network has been proposed, where carrier modulated signal is introduced on top of the existing CAN signal whereby the data rate can be enhanced over 100Mbps. The proposed scheme is compatible with the existing CAN network and accordingly enables seamless upgrade of the existing network to support high speed demand using CAN protocol.*

## Introduction

With the exploding needs for smarter cars, the needs for higher data transmission links within the car are rapidly growing. CAN, LIN and FlexRay are among the standards that are used for in-vehicle link solutions [1]. CAN has been the preferred choice due to its low complexity using bus topology, real time response and robustness to errors, resulting in the wide spread adoption across all types of vehicles and applications such as driver protection assistance systems [2]. With many electronic control units (ECU) and passive nodes connected on the same bus, the priority-based fast contention resolution of CAN standard is best suited for safety critical applications. On a flip side, however, nodes with lower priority messages can take very long delays to be able to get a chance to transmit. The delay characteristics of message transmission in real environment attracted a lot of research efforts and there have been numerous researches done using both analytic and model-based methods [3~5]. As a solution to this delay issue, a modified CAN standard called time triggered CAN (TTCAN) standardized in ISO 11898-4 was proposed, which is a higher layer protocol operating on top of CAN [6]. In TTCAN, there are windows in time domain during which only one node is allowed to transmit. Also there are other windows during which all nodes can compete for transmission. Hence, nodes with lower priority messages can use windows dedicated to them for data transmission without suffering severe delay.

Another aspect of improvement on the CAN system is related to the improvement of its transmission rate. Most widely adopted CAN system now supports maximum data rate of up to 1Mbps, which is shared between all the devices connected to the same bus. Considering the data rate requirement for the application of CAN to bandwidth demanding applications like rear seat entertainment (RSE) and driver assistance system using vehicle cameras, its data rate still falls far short of the requirements, compared to what other schemes offer like Ethernet and Fiber system offer. There has been lot of efforts within CAN eco-system to improve the data rate beyond 1Mbps. As one of the most obvious improvements, overclocking of the CAN signal has been proposed to increase the data rate to over 1Mbps [7~9]. In [7], overclocking of the data field has been proposed, by which data field in a CAN frame can be transmitted at a higher clock rate reducing the time spent for data field transmission. However, the overall data rate improvement quickly reaches certain limit because the maximum bit in data field is limited to 64 bits. CAN-FD (flexible data rate) has been proposed as an upgrade to CAN 2.0 where both overclocking and extension of data field are introduced for higher data rate [8]. The data rate is increased by sending bits in data field at higher clocking rate and by introducing longer data field length. However, due to the overclocking during the data field transmission, CAN-FD is not compatible with existing CAN standard devices that can observe bit transitions within a CAN bit period, subsequently reporting errors. As a solution to avoid this unwanted error condition during the CAN bit period, overclocking only in the part of a duration that is not observed by other CAN devices on the bus is introduced and shown to support higher data transmission up to 16Mbps [9]. Even with these improvement, the data rate is still lower than what other standards support to meet the increasing demand for higher data rate. This paper proposes a new scheme to increase the data rate of CAN system to provide data rate over 100Mbps while keeping compatibility with CAN standard.

The paper is organized as follows. Section 2 briefly introduces the CAN standard and protocol. Section 3 describes high-speed CAN signal transmission scheme and presents transmitter and receiver to implement the scheme. Section 4 provides the simulation results for the proposed scheme. In section 5, further works for the implementation is discussed. Section 6 concludes the paper.

## Controller Area Network (CAN) System

Conventional CAN system is based on the bus configuration as in Figure 1. All the nodes share a bus for communication and each node can transfer data to any node on the bus. Each node comprises a controller and a transceiver. The controller performs bit transmission/reception and timing control. CAN transceiver is responsible for signal conversion to and from the bus. In the typical application, the two lines on the bus denoted by CAN_H and CAN_L swings from 2.5V in opposite direction with offset of 1V, producing differential signal when transmitting dominant bit equal to 0. When transmitting recessive bit equal to 1, the transceiver stops driving the bus and the levels of the two lines return back to 2.5V. The resulting CAN signal is transferred on the CAN bus as in Figure 2 in the form of differential signal to reduce the interference from other noise sources. The slew rate of level change from dominant to recessive and vice versa can be controlled to control the emission on the bus. The bus is terminated on both ends to prevent signal reflection from degrading the quality of the received signal on each node. On the receiving node, a CAN receiver detects the voltage level difference between CAN_H and CAN_L to determine bits that are being transferred.

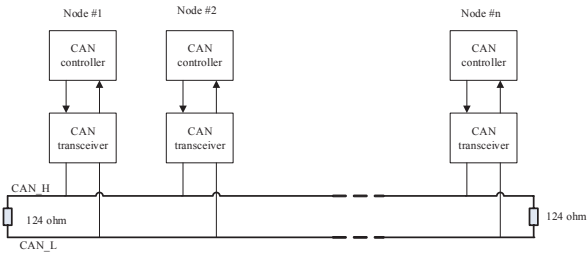

*Figure 1. CAN bus configuration*

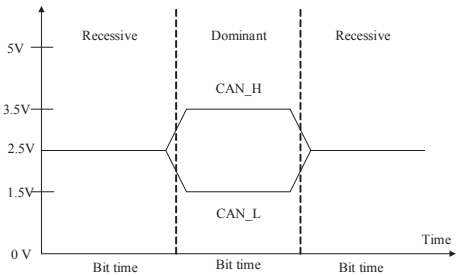

*Figure 2. Dominant and Recessive bits*

Data in the CAN system is transferred in the form of frame as shown in Figure 3. Let dominant bit and recessive bit be denoted by D and R, respectively. In idle period, the bus stays at R with no node driving the bus. The transmitting node attempts to send the frame by transmitting D when the bus is confirmed to be in the state of R. The initial transmission of D is called start-of-frame (SOF). Then, arbitration field made up of 12 bits follows SOF. The arbitration field contains 11 bit field that represents unique identifier indicating the type of message it wants to send. In the case when two nodes are starting SOF and transmitting identifiers at the same time, the scheduling is performed based on priority of the identifiers, using so-called carrier sense multiple access with bitwise arbitration (CSMA/BA). The node with smaller value in identifier value prevails and wins the exclusive right to continue to drive the bus. The losing nodes stops transmission and switches to listening state, waiting for next chance. This unique and fast resolution of contention is one of the greatest benefits that CAN provides compared to other standards such as Ethernet where conflict of transmission is resolved by each of the transmitting node transmitting after random periods with none of the nodes getting message transmitted immediately. This fast contention feature is not only useful to real time control purpose but also very effective in enhancing overall bus throughput of the CAN system.

Control field contains controls bits and bits indicating the number of data following it. In the data field, up to 8 byte or 64 bits can be transferred in CAN 2.0. In this case, CAN frame consists of 110 bits for standard format. Since CAN standard specifies that every 5 consecutive transmission of same bits, either of D's or R's, should be followed by a transmission of a different bit that is inserted at the transmitter, transmitted on the CAN bus and removed from the receiver. Hence, the actual number of bits transmitted on the CAN bus could be larger than 110 bits due to this dummy bit insertion and removal. In CAN-FD, the duration of data field can be increased over 64 bit periods specified in CAN 2.0. CRC field is used to check data integrity. Receiving nodes calculate CRC of the frame and compares it against the received CRC. ACK field is used to inform the transmitter of the receipt of the frame. In the case of error, receiving nodes can start transmission of error frame to report the error. Following ACK are end-of-frame (EOF) of 7 bits and inter-frame-space (IFS) of 3 bits. After IFS, any node can start transmission by sending SOF.

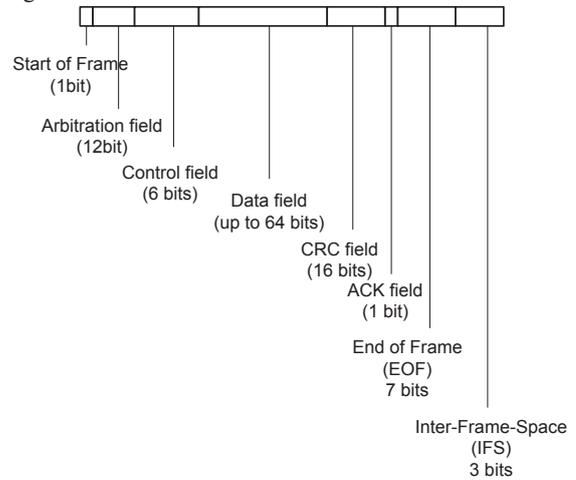

*Figure 3. CAN frame*

The data rate of CAN is directly proportional to the clocking rate of the bits. However, the clocking rate of bits cannot be arbitrarily set to high values, because higher rate pulse with shorter bit period on the CAN bus can suffer from sever degradation in signal integrity resulting in loss of data in the receiver due to failure of edge detection and reduction in the level of signal. In CAN-FD where up to 16 times of the original clock is considered, the use of higher clocking pulses is dependent upon the length of the bus and termination condition. CAN-FD can be considered as an upgrade to

CAN for purpose of supporting incremental increase in bandwidth by limited overclocking.

## The Proposed High Data-Rate CAN System

The primary cause of the data rate limitation of the CAN system comes from three factors. One is the constraint of the bus characteristics that limits the minimum clock pulse width which then limits maximum clock rate. Secondly, due to the attenuation in higher frequency, higher clock pulse suffers from severe edge degradation that could render received waveform hard to detect properly. Finally, only binary signaling is allowed in the standard with very low bandwidth utilization.

The proposed scheme overcomes these limitations by adding carrier modulation to the CAN frame along with higher bandwidth utilization. One of the biggest advantage of using carrier modulation for data transmission is that the proposed system is no longer dependent upon edge detection using the bit transitions. It also enables the use of higher bandwidth modulation sending multiple bits for each transmit symbol providing higher data rate without transmission bandwidth increase.

### Introduction of carrier modulated signal

The proposed scheme applies carrier modulated signal on top of the standard CAN signal, when dominant bits are transmitted by a transmitting node. Figure 4 shows the case for the propose scheme when there are three nodes connected on the bus. The proposed CAN node is transmitting on the bus while there are two nodes receiving from the bus, one being standard CAN node and the other being high-speed CAN node. The proposed CAN transmit node is composed of controller and transmitter. The proposed high-speed CAN transmitter is designed such that the transmitted high-speed CAN signal does not cause any error condition to the existing standard CAN receiver. On the other hand, high-speed CAN receiver conforming to the proposed scheme can recognize the carrier modulated signal and perform required demodulation. The proposed high-speed CAN controller sends both high speed CAN transmit bits and standard CAN transmit bits to the transmitter. The standard CAN transmit bits are comprised of the standard CAN frame as in Figure 3. High-speed CAN transmit bits are split into two streams for in-phase and quadrature modulation. Depending upon the modulation scheme used, the bits in each stream is grouped into $N/2$ bits, where $N$ is an even integer. So for each in-phase/quadrature symbol, $N$ bits are transmitted. It can be seen that $N=2$ and $N=4$ corresponds to the modulation scheme of quadrature phase shift keying (QPSK) and 16 quadrature amplitude modulation (QAM), respectively. The pulse shaping on complex in-phase/quadrature symbol can be additionally applied for band limiting purpose. Then, carrier modulation is applied to get the carrier modulated CAN signal, which is applied on top of the standard CAN signal. Then CAN bus signal generator converts the signal for transmission on the bus.

Figure 4 depicts the high-speed CAN signal generator, where carrier modulated signal $S_p(t)$ is first scaled by $A_p$ and then shifted by fixed offset of typical 1V during dominant bit transmission. The resulting high-speed CAN signal $Q_s(t)$ is applied to the CAN bus signal generator only when dominant bits are transmitted in the CAN frame. $Q_s(t)$ is then sent to the bus signal converter to produce differential signal $Q_d(t)$ to be transmitted on the bus. It should be noted that the value of $A_p$ determines the output power of the carrier modulated signal with respect to standard CAN signal. Higher value of $A_p$ is preferred for reliable transmission of carrier modulated CAN signal. However, setting $A_p$ too high can cause $Q_s(t)$ to swing below the threshold level of dominant bit detection on the standard CAN receiver. Hence, $A_p$ should be set to proper value depending upon bus configuration and system bit-error-rate requirements. For instance, $A_p$ could be simply set such that the lowest value of $Q_s(t)$ during dominant period should be higher than 0.5V for the purpose of preventing other receivers from erroneously detecting recessive bits. Other sophisticated scheme of $A_p$ can be also considered using the power of $Q_s(t)$.

In idle period, the proposed high-speed CAN controller starts frame transmission in the same way as the standard CAN controller by sending SOF and then arbitration field to gain bus access. If the node wins the bus through standard arbitration process, high-speed CAN signal transmission can start from CAN data field. Note that the CAN bits are transmitted as the standard at the maximum rate of 1Mbps while high speed bits are transmitted on the modulated carrier with higher data rate. When the dominant bit, D, is to be transmitted, the combined signal is the carrier modulated signal shifted by a fixed offset. During recessive bit transmission no signal is transmitted, because the CAN transmitter is allowed to drive the bus only when it is transmitting dominant bits of CAN. When sending R bits, both generation of carrier modulated signal $S_p(t)$ and transmission of high-speed CAN signal stop. It can be easily seen that the period of carrier modulated signal is proportional to the number of dominant bits in the data field of the standard CAN frame. The transmit period of high-speed CAN signal is maximized when the size the data field is set to the maximum of 64bits. Further increase in data rate can be achieved when all the data bits are set to dominant 0, enabling the transmission of high-speed CAN signal over the whole data field, resulting in highest data throughput. We will consider this case only in this paper, while it is true that sending meaningful data in the standard CAN frame at the cost of reduction in overall throughput can bring benefits depending upon application. It should be noted that CAN bit stuffing rule dictates to put a bit of opposite polarity after every 5 consecutive bits, while the stuffed bits are removed on the receiver side. Hence, setting data field bits to all D's will result in the bit pattern of transmission of 5 dominant bits followed by transmission of one recessive bit, which repeats itself during the data field transmission, i.e., repetition of the pattern (D-D-D-D-D-R).

### Introduction of Multi-level Modulation Scheme

In contrast to the binary signaling of CAN system, the use of multi-level modulation is considered to increase the throughput. High-speeds CAN transmit bits are used to form complex symbol to be modulated and transmitted on the bus. The complex symbol can be constructed using any modulation scheme such as QPSK, 8-PSK and 16QAM, 64 QAM etc. Higher order modulation is preferred to get higher data rate. However, the choice of modulation scheme is related to the signal-to-noise ratio, frequency attenuation characteristics of the channel and receiver complexity.

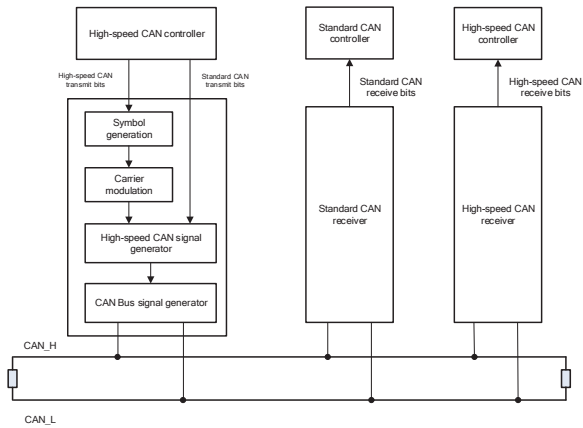

Figure 4. Proposed high-speed transmitter

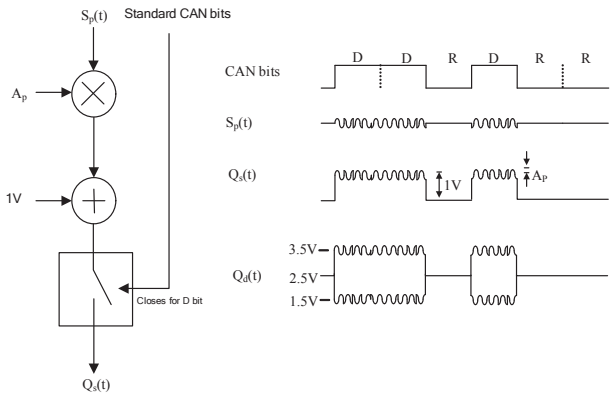

Figure 5. High-speed CAN signal generation

### Carrier frequency selection

As for carrier frequency selection of modulated signal $S_p(t)$, using lower carrier frequency can be beneficial since attenuation is lower in lower frequency region in CAN bus environment. However, this can cause high frequency noise generated from the transition of the standard CAN signal to interfere with $S_p(t)$ especially in the start and end part of high-speed CAN signal. Setting carrier frequency higher will reduce interference but incur higher attenuation of modulated signal. Carrier frequency is a parameter to be optimized according to the channel characteristics and modulation schemes.

### Receiver

Figure 6 shows the proposed receiver. The differential signal received from the bus is converted to single-ended signal, using differential-to-single conversion device, which is applied to standard CAN bit detector and band-pass filter for high-speed signal demodulation, respectively. Standard CAN detector monitors the bus and detects the start of recessive-to-dominant bit transition during data field and enables the high-speed CAN demodulator operation to run for the period of 5 D bits. Bandpass filter in the high-speed CAN receiver removes the interference from standard CAN signal as well as out-of-band noise, providing input to the demodulator and equalizer of the proposed scheme. The demodulator down-converts carrier modulated signal to generate baseband signal providing in-phase and quadrature components. Output of demodulator is applied to equalizer for channel compensation. For the training of the equalizer, simple LMS (least mean square) algorithm is considered [10].

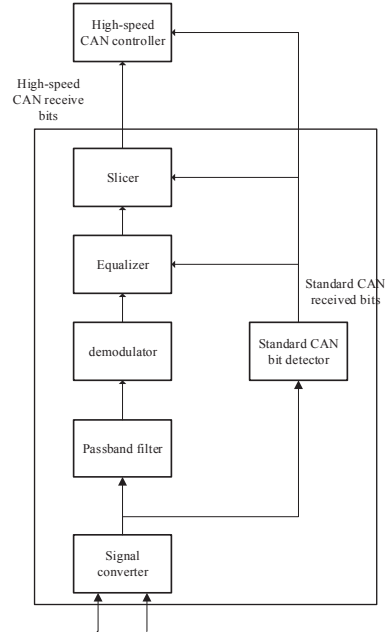

Figure 6. Proposed receiver

## Simulation Results

The receiver of the proposed scheme uses the bit pattern of the data field to run the receiver for the period of 5 bit durations and hold the operation for one bit period. This process is repeated until the end of data field. Figure 7 shows the simulation setup. Standard CAN frame with data field bits set to all D's is generated to make CAN signal while random bit generators are used to generate bits to be carried in $S_p(t)$. The combined signal $Q_d(t)$ passes dispersive channel experiencing frequency-dependent distortion. Additive white Gaussian noise (AWGN) is added before the signal is input to receiver for demodulation. Carrier frequency of $S_p(t)$ is set to 24MHz and symbol rate of $S_p(t)$ is 36MHz. 16QAM was used for symbol generation. For the purpose of simplicity, the carrier frequency and timing of both transmitter and receiver is assumed to be perfectly matched.

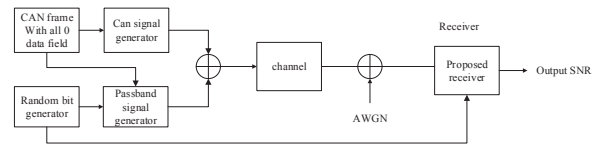

Figure 7. Simulation setup

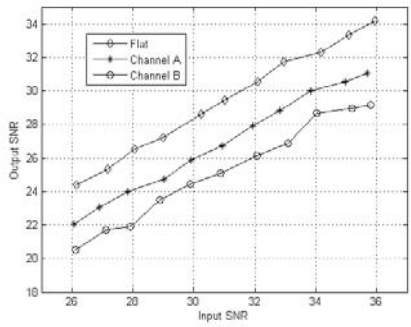

*Figure 8. Equalizer output SNR vs Input SNR*

The equalizer in the receiver employs symbol spaced equalizer for feed-forward part with 24 taps and 8 taps are used in feedback filter [10]. In order to train the equalizer in the receiver for each CAN frame, QPSK training symbols made of random bits are sent for the period of the first 15 dominant bits in the data field. The receiver uses conventional LMS method to train its coefficients with center tap is initialized. Once training is complete, modulation scheme is switched to 16QAM where the equalizer uses decision directed training until the end of data field.

Two channels for the CAN bus are used for performance evaluation. A channel denoted by Channel A is derived from the worst-case UTP model in [11]. Channel A corresponds to the length of 100m with CAT-3 cables. In real car environment, the length of the bus can be much shorter than this. So channel A can be considered as worst case end-to-end channel distortion on the CAN bus. Another channel denoted by Channel B is a modification of channel A by inserting 9 CAN nodes of input impedance of 20K ohm in the middle each 10 meters apart, where the length of the feed-line on each node is set to 30 cm.

The transmitter is designed to send CAN frames repetitively for each simulation setup of input SNR. The receiver performs equalizer training for each CAN frame independently. On the receiver side, the output SNR of the equalizer is calculated comparing the equalizer output against transmitted symbol and used as performance measure. Figure 8 shows the performance of the proposed scheme for varying channel SNR in the case of 3 different channels. Horizontal axis represents SNR at the input of the receiver. For an ideal channel without frequency distortion denoted by Flat, the SNR loss of 1.5dB at the output of equalizer is observed and this is due to the loss caused by interference from CAN signal and imperfect equalizer training. For channel A and channel B, compared to flat channel case, higher loss in SNR can be seen due to the frequency attenuation and resulting equalizer loss. Channel A show loss of about 4dB resulting from the attenuation in the high frequency. Channel B results show loss of about 6 dB because the CAN nodes in the middle causes even higher attenuation in high frequency compared to channel A. However, it can be seen that the proposed scheme can provide high rate communication links, e.g., 16QAM, with manageable SNR loss even by employing simple conventional LMS equalizer. More sophisticated equalizer scheme can be used to reduce the SNR loss.

Figure 9 shows the net bit rate ratio, defined by the net data rate divided by maximum data rate according to the data field size. The overhead bits includes SOF, arbitration field, control field, CRC, ACK, IFS, equalizer training bits and stuffing bits. It can be seen that increasing the data field beyond 64 bit period of CAN standard can significantly improves the overall throughput reducing the effects of overhead bits. With symbol rate equal to 36MHz and 16QAM used, the data rate is 128Mbps. If data field length of 1024 bits or more can be used as in the case of CAN-FD to get net data rate ratio of about 0.8, more than 100Mbps can be achieved.

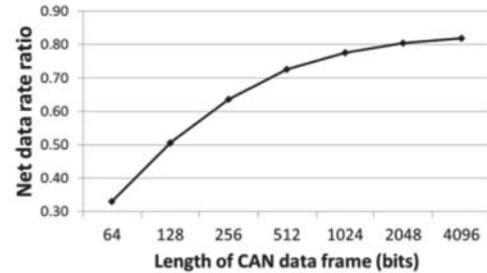

*Figure 9. Net data rate ratio according to the length of data field*

### Further Works

For the equalizer, the proposed scheme used simple LMS training with long training sequence. However, more efficient equalizer initialization scheme can be used to reduce the length of training bits in the data field in order to increase the efficiency and performance. Finally, the frequency and timing of the transmitter and receiver was assumed to be synchronous. The introduction of appropriate synchronization signal for timing and carrier acquisition/tracking prior to the data transmission will be required for real implementation.

### Conclusions

A new scheme for improving the data rate of CAN system has been proposed by introducing carrier modulation on top of the existing CAN signal. The performance of the scheme in CAN bus environment has been evaluated to show that the proposed scheme can provide higher data rate while keeping backward compatibility with the CAN standard. With simple modification of the CAN frame to extend data field, the net data rate can be increased to over 100Mbps.

The inherent advantage of bus topology supported by CAN standard can be very beneficial to save the amount of wiring required to interconnect all the devices compared to star topology. This will help reduce the cost and weight of the vehicle. By enabling higher data transmission with the proposed scheme, CAN standard will be able to expand its application to versatile in-vehicle devices requiring higher bandwidth beyond its wide adoption to control devices.

**Acknowledgements**

This work was partly supported by ICT R&D program of MSIP/IITP [14-824-09-013, Resilient Cyber-Physical Systems Research] and DGIST R&D Program of MSIP (14-BD-0404).

*Short bio (75 - 100 words) per author.*

Suwon Kang received the B.S., M.S., and Ph.D. degrees in electrical engineering from Seoul National University (SNU), Seoul, Korea, in 1993, 1995, and 2001, respectively. His research interests are wireless and wire-line communication systems and signal processing

Sungmin Han received the B.S. degree in Electronics Engineering from Korea University of Technology and Education, Cheonan, Korea, in 2012. Currently, he is working toward his PhD degree in Department of Information and Communication Engineering (ICE), Daegu Gyeongbuk Institute of Science and Technology (DGIST), Daegu, Korea. His research areas are communication theory and communication networks.

Seungik Cho received the B.S. degree in Information and Communication Engineering from Chungbuk National University (CBNU), Cheongju, Korea, in 2014. Since 2014, he has been studying in Department of ICE, DGIST, Daegu, Korea as a master student. His research area is advanced communications and RF energy harvesting.

Donghyuk Jang received B.S. degree in Electronics engineering from Kyungpook National University (KNU), Daegu, Korea, in 2013. He joined DGIST, Daegu, Korea, in 2013, where he is currently an MS candidate. His research interest is channel modeling and analysis of CAN communication system.

Hyuk Choi received the B.S., M.S., and Ph.D. degrees in electrical engineering from SNU, Seoul, Korea, in 1994, 1996, and 2002, respectively. In 2003, he joined the faculty at University of Seoul, where he is currently a Professor in the School of Computer Science. His research interests are in information security and signal processing.

Ji-Woong Choi received B.S., M.S. and Ph.D. degrees in Electrical Engineering from SNU, Seoul, Korea, in 1998, 2000 and 2004, respectively. From 2004 to 2005, he worked as a Postdoctor in ISRC, SNU. From 2005 to 2007, he was in Stanford University, USA, as a Postdoctor. From 2007 to 2010, he worked for Marvell Semiconductor, USA, as a Staff Systems Engineer for WiMAX and LTE system design. In 2010, he joined DGIST, Daegu, Korea, where he is currently an Associate Professor. His research areas are communication theory and signal processing.